\documentclass[aps,prl,twocolumn,showpacs,superscriptaddress]{revtex4}
\usepackage{graphicx}
\usepackage{amssymb}

\newcommand{\beq}{\begin{equation}}
\newcommand{\eeq}{\end{equation}}

\begin{document}

\title{Scaling law for seismic hazard after a main shock}

\author{Stefano Lise}
\affiliation{Mathematical Physics, Department of Mathematics,  
Imperial College London, London UK SW7 2AZ.}
\affiliation{Department of Biochemistry \& Molecular Biology, 
    University College London, Gower Street, London WC1E 6BT, UK}
\author{Maya Paczuski} 
\email{maya@ic.ac.uk}
\affiliation{Mathematical Physics, Department of Mathematics,  
Imperial College London, London UK SW7 2AZ.}
\author{Attilio L. Stella}
\affiliation{INFM-Dipartimento di Fisica, Universit\`a di Padova, 
I-35131 Padova, Italy.}

\begin{abstract}
{After a large earthquake, the likelihood of successive strong
aftershocks needs to be estimated. Exploiting similarities with
critical phenomena, we introduce a scaling law for the decay in time
following a main shock of the expected number of aftershocks greater
than a certain magnitude. Empirical results that support our
scaling hypothesis are obtained from analyzing the record of
earthquakes in California.  The proposed form unifies the well-known
Omori and Gutenberg-Richter laws of seismicity, together with other
phenomenological observations. Our results substantially modify
presently employed estimates and may lead to an improved assessment of
seismic hazard after a large earthquake.} 

\end{abstract}


\maketitle

\begin{figure}[hb]
\begin{center}
\includegraphics[width=90mm]{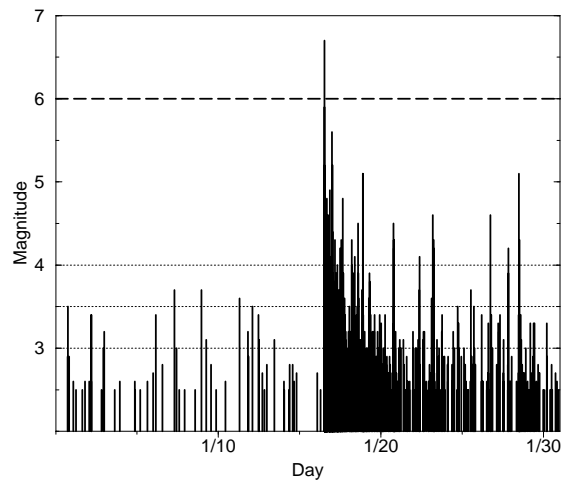}
\end{center}
\caption{
\label{sc_ts}
Time series of earthquakes in Southern California 
in the month of January $1994$. The Northridge earthquake of magnitude
$6.7$ occurred on January 17. The dashed line represents a main shock
threshold $M=6$ and the dotted lines represent aftershock thresholds 
$m=3$, $3.5$ and $4$.
}
\end{figure}

Earthquakes tend to occur in clusters~\cite{scholz_2}. A large
earthquake often precedes many further aftershocks, some of which may
themselves be consequential.  For instance, the Northridge earthquake
in Southern California set off markedly increased seismic activity in
the area, as apparent in the time series shown in
Figure~\ref{sc_ts}. In populated regions like Southern California that
are prone to large magnitude events, the likelihood of occurrence of
strong aftershocks requires accurate assessment in order to determine
the risk of further damages~\cite{reasenberg_1,reasenberg_2,jap_gov}.

Earthquakes arise from a complicated nonlinear 
mechanics~\cite{scholz_2,huang,scholz_1} and individual events are 
not  predictable~\cite{geller,nat_deb}. Yet, simple scaling laws 
describe their statistical properties~\cite{scholz_2,kagan,bak_02}. 
Empirically, the number $n(t)$ of aftershocks occurring at time $t$ after a 
large magnitude event obeys the modified Omori
law~\cite{omori,utsu_1,utsu_2}
\begin{equation}
\label{omori_law}
n(t) = \frac{K}{(c+t)^p} \qquad .
\end{equation}
The constants $K$ and $c$ are positive, and the exponent $p$ is
usually found to be close to one. The Gutenberg-Richter (GR)
distribution~\cite{gr} for the frequency of earthquakes with magnitude
greater than $m$ in a seismic region is
\begin{equation}
 \label{gr_law}
 {\cal N}(m)= A \: 10^{- b m} \qquad ,
\end{equation}
where $A$ is a constant and the critical exponent $b$ typically takes a 
value close to one.  Since the magnitude is proportional to the logarithm of 
the size of the earthquake, Eq.~(\ref{gr_law}) describes a scale-free 
distribution.

Comparisons between seismicity and critical phenomena in other physical 
systems~\cite{bak_02,bak,turcotte} suggest that these laws, usually 
considered to be independent, should connect inextricably
together. Progress toward an organizing theory of seismicity could
therefore emerge along lines similar to those followed in the modern
approach to critical phenomena~\cite{kadanoff}. Here, we show that, in
fact, the rate of earthquakes following a main shock obeys a unified
scaling law. This law relates the expected number of aftershocks greater 
than a certain magnitude to the magnitude of the main shock itself, and to 
the time since the main shock. It interpolates between the Omori law at
short times and the GR relation at late times, providing a coherent
framework from which several other phenomenological observations can be 
deduced.  In addition, the mathematical form for the rate of aftershocks 
differs significantly from the presently employed formula used to assess 
seismic hazard~\cite{reasenberg_1,reasenberg_2,jap_gov}.

Analysis of earthquake recurrence times use a magnitude threshold above 
which events are registered~\cite{utsu_2}. This magnitude threshold always 
exists, even if not explicitly specified, due to limitations in reliably 
detecting all small earthquakes in a seismic region. Recently, Bak {\it et 
al} considered the magnitude threshold itself as a variable entering into a 
scaling law for waiting times between subsequent events~\cite{bak_02}.
As we show below,  threshold variables also enter into a scaling theory
for rates of aftershocks.

In order to quantify increased seismicity following a large earthquake, we
separate smaller events from larger ones  by imposing two different
magnitude thresholds, as indicated in Figure~\ref{sc_ts}. The 
threshold $M$ defines the large events.  Earthquakes larger than magnitude
$M$ are referred to as main shocks.  A smaller threshold,
$m$, selects the remaining events in the catalog to be counted.
These remaining events, larger than magnitude $m$, are referred to
as aftershocks.  The quantity $R_{M,m}(t) dt $ is the average number
of events in the catalog with magnitude larger than $m$ occurring in
the interval $(t,t+dt)$, given that an isolated main shock of magnitude 
larger than $M$ occurred at $t=0$.

We analyzed the earthquake catalog of Southern
California~\cite{cal_cat} from 1984 to 2002 in order to measure the
rate $R_{M,m}(t)$.  This catalog includes
more than $3.5 \times 10^5$ earthquakes.  We identified all events
larger than magnitude $M$ ($5.5 \le M \le 6.5$). For each main shock,
we measured, as a function of time, the rate of subsequent events
larger than magnitude $m$ with $2.5 \le m \le M-1.5$.  The resulting
data therefore contain background seismicity, unlike some other
investigations~\cite{kisslinger}.  The measurement of the rate following a 
given main shock was stopped when another event of magnitude $M$ or larger 
occurred. This ensures that each aftershock sequence does not include events 
from other main shocks. Some aftershock sequences, therefore, comprise more 
events and last longer than others.  To avoid the effects of correlations 
between main shocks, we considered only isolated ones. We excluded 
all events that follow the second of two successive main shocks, where 
both main shocks occur within a time interval 
(of the order of $4$ months) of each other. An average over main shocks was 
then computed for each pair $(M,m)$ to obtain $R_{M,m}(t)$.

Depending on the value of $M$, the same earthquake becomes
a main shock in one rate measurement,  and  an aftershock in another. 
For example, a magnitude $5.8$ earthquake could be an aftershock of a
magnitude $6.7$ earthquake if $M=6$, but is treated as a main shock 
if $M=5.5$. This approach reflects the view that aftershocks do not differ 
from other earthquakes and can therefore lead to aftershocks 
themselves~\cite{hough}.  In this way, the pattern of seismicity in space 
and time can be considered to be a single, hierarchically organized, 
critical process~\cite{bak_02,bak}.

\begin{figure}[hb]
\begin{center}
\includegraphics[width=90mm]{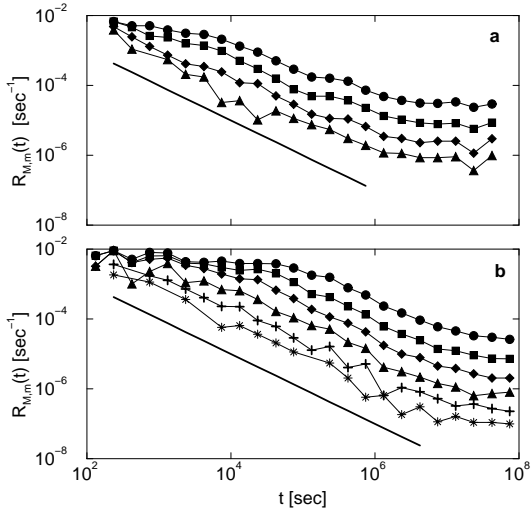}
\end{center}
\caption{
\label{rate}
Average rate of aftershocks at time $t$ after a main shock in Southern 
California. In panel (a) the 
main shock has magnitude greater than $M=5.5$, in panel (b)  greater than 
$M=6.5$. Averages are made over (a) $29$ and (b) $4$
main shocks. Symbols corresponds to different threshold magnitudes for
aftershocks, respectively
($\bullet$)~$m=2.5$,
($\blacksquare$)~$m=3$,
($\blacklozenge$)~$m=3.5$,
($\blacktriangle$)~$m=4$,
(+)~$m=4.5$ and
(*)~$m=5$.
The exponent of the straight line is $-1$.
}
\end{figure}

Figure~\ref{rate} shows the empirical results. At short times
the rate, $R_{M,m}(t)$, approaches a universal constant value, 
$R_0 \simeq 10^{-2}$, consistent with the presence of the parameter $c$ 
in Omori's Law, Eq.~(\ref{omori_law}). Strikingly, the rate at short times, 
$R_0$, appears to be independent of both $M$ and $m$, indicating that soon 
after a main shock earthquakes of all magnitudes are equally probable.
This observation contrasts with the rate estimate used in
Refs.~\cite{reasenberg_1,reasenberg_2,jap_gov}, where the rate at short
times depends on both thresholds $M$ and $m$. However, the constant regime 
persists longer for increasing $M$ and/or for decreasing $m$.  Later in
time, a $1/t$ power law decay sets in.  Each rate eventually reaches a 
stationary value at late times.  The stationary rates for different 
thresholds $m$ agree with  the GR law, and do not depend on $M$. The
duration of aftershock activity, $t_d$, is defined as the time for the 
aftershock rate,  $R_{M,m}(t)$, to decline to its stationary level. This 
level is referred to as background seismicity. Comparing Figs.~\ref{rate}a 
and~\ref{rate}b shows that the duration of aftershock activity increases as 
the magnitude threshold of the main shock $M$ increases.

All these observations can be combined into a single  scaling hypothesis for
$R_{M,m}(t)$
\begin{equation}
\label{scaling_law}
R_{M,m}(t) = 
\left( \frac{1}{t_0 + 10^{b(m-M)} \, t} \right) G(A \, 10^{- b M} \, t) 
\qquad .
\end{equation}
For $R_{M,m}(t)$ to approach a stationary value at late times, the scaling 
function $G(x)$ must behave as $G(x) \propto x$ for large $x$.  In that 
case the stationary rate at late time
is proportional to $10^{-b m}$, consistent with the GR law.
At early times, $G(x)$ for small $x$ is evaluated.  If this is constant,
the first term on the 
right hand side of Eq.~(\ref{scaling_law}) controls the behavior.  This
term corresponds
to an  Omori law 
with $p=1$.  The parameter $A$ and the critical exponent $b$ were determined
from the GR relation, Eq.~(\ref{gr_law}), to be 
$A = 10^{-2 \pm 0.2} \, \mbox{sec}^{-1}$ and $b=0.95 \pm 0.1$ for the
Southern California data set. The only remaining parameter in 
Eq.~(\ref{scaling_law}) is the time  $t_0$.

\begin{figure}[hb]
\begin{center}
\includegraphics[width=90mm]{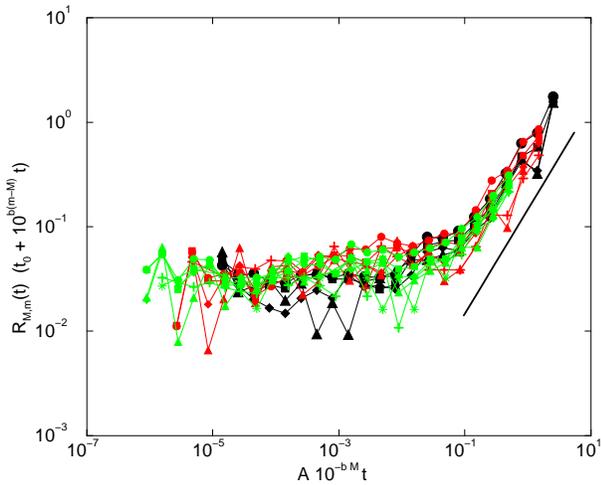}
\end{center}
\caption{
\label{rate_s}
Data collapse for the rate of aftershocks following a main shock in 
Southern California.  This defines the crossover function $G(x)$.
 The numerical values of the parameters are $b=0.95$, 
$A=10^{-2} \: \mbox{sec}^{-1}$ and $t_0=6 \: \mbox{sec}$. Note that only 
$t_0$ is a fitting parameter, since the others are determined from the
GR relation. Symbols are as in fig.~2, colored in black, 
red and green respectively for main shocks of magnitude greater than  
$M=5.5$, $M=6.0$ and $M=6.5$. The exponent of the straight line is $1$.
}
\end{figure}

The data collapse method can be used to test and potentially falsify 
Eq.~\ref{scaling_law}. 
This is accomplished by rescaling the time, $t$ by 
$\left(A \, 10^{-b M}\right)$ and the 
rate, $R_{M,m}(t)$ by $\left( t_0 + 10^{b(m-M)} \,t \right)$, to get a 
dimensionless scaling plot for $G(x)$. Using 
$t_0=6 \, \mbox{sec}$, we find that all the curves collapse onto
a single curve, as shown in
Figure~\ref{rate_s}, within statistical error.  The data collapse verifies
our scaling hypothesis.
 The value of $t_0$ has an uncertainty; we estimate
$4 \, \mbox{sec} \le t_0 \le 12 \, \mbox{sec}$ from changes in the 
quality of the data collapse using different values for $t_0$. 
Evidently,  $G(x)$ changes behavior at a turning point $x=x_0$.
Two regimes are distinguished: a transient Omori limit at short times, 
$G(x \ll x_0) \rightarrow const.$, and a stationary GR limit 
at late times,  $G(x \gg x_0) \rightarrow x$.

According to our scaling hypothesis, the duration of correlated aftershock 
activity depends on the threshold  magnitude, $M$, of the main shock as
\begin{equation}
\label{duration}
t_d \simeq \frac{x_0}{A} \, 10^{b M} \qquad .
\end{equation} 
As $x_0 \simeq 10^{-1}$ (see Figure~\ref{rate_s})  this implies, for example,
that, on average, the excess rate of aftershocks persists for about two months
following an earthquake of magnitude larger than six.

The parameter $c$ in Omori's law is a controversial quantity
in the earthquake literature~\cite{utsu_2}. Reported values range from less 
than 
$0.01 \, \mbox{days}$ to over $1 \, \mbox{day}$. Previous investigations 
usually analyzed aftershock sequences following specific  main shocks of 
interest. It  was found that the value of $c$  decreases with increasing 
threshold magnitude $m$ of aftershocks included in the analysis.
Comparing Eq.~(\ref{omori_law}) with the scaling  hypothesis 
(\ref{scaling_law}), we obtain 
\begin{equation}
\label{c_value}
c= t_0 \, 10^{b(M-m)} \qquad .
\end{equation}
This sheds light on the variability of the $c$-value found in previous 
investigations.

The value of the time, $t_0$, can be related to the dimensionless variable, 
$x_0$, and the universal short
time rate, $R_0$. Indeed the $1/t$
decay of the rate  at intermediate times implies that 
$R_0 \, c \simeq R_{M,m}(t_d) \, t_d$. As 
$ R_{M,m}(t_d) \simeq A \, 10^{-b m}$
and from expressions~(\ref{duration}) and~(\ref{c_value}), it follows that
\begin{equation}
t_0 \simeq \frac{x_0}{R_0} \simeq 10 \, \mbox{sec} \qquad ,
\end{equation}
which is within the range of uncertainty previously estimated.
This provides a consistency check on our statistical analysis.
Conversely, the scaling behavior, Eq.~ (\ref{c_value}), of the $c$-value 
arises from the condition that the rate at short times $R_0$ is 
approximately independent of $M$ and $m$.


After a large earthquake, forecasts of seismic hazard provided by
government agencies are based
on estimates of the rate $\lambda _{M,m} (t)$ of 
aftershocks with magnitude greater than $m$ occurring at time $t$ after a
main shock of magnitude $M$~\cite{reasenberg_1,reasenberg_2,jap_gov}.
Within our ansatz, the rate  $\lambda _{M,m} (t)$ is
\begin{equation}
\label{exp_rate}
\lambda _{M,m} (t) = R _{M,m} (t) - R _{m} (\infty)- \frac{1}{b \ln 10} 
\frac{\partial R _{M,m} (t)}{\partial M}
\end{equation}
where $R _{m} (\infty)$ is the background seismicity, generally
not included in aftershock probability evaluation. Eq.~(\ref{exp_rate}) is
only valid when $m<M$. It is important to underline that 
$\lambda _{M,m} (t)$ as derived here in Eq.~(\ref{exp_rate}), from an 
empirically tested law, Eq.~(3), differs  in several respects from
presently employed estimates~\cite{reasenberg_1,reasenberg_2,jap_gov}.
For example, it predicts that the rate of aftershocks at early times does
not depend on the magnitude of the main shock or on the aftershock
threshold (as discussed above, in this limit 
$\frac{\partial R _{M,m} (t)}{\partial M} \simeq 0$). On the contrary, in 
the stochastic model employed in~\cite{reasenberg_1,reasenberg_2}, 
$\lambda _{M,m} (t) \sim 10^{b(M-m)}$ immediately after the mainshock.


In conclusion, we have introduced and tested against the record of recent 
earthquakes in California, a new scaling law that unifies the Omori law for
aftershocks and the 
GR relation, previously considered to be independent.
Our finding indicates a theoretical framework
(scaling theory) within which the problem of earthquake occurrence should be
considered. It provides an empirical characterization of earthquake 
statistics that will be useful to test theories and models in the
future. Our results may also have practical implications, namely
improved hazard assessment following a large earthquake.

\end{document}